\documentclass[conference,a4paper]{APSIPA2025}
\usepackage{amsmath}
\usepackage{graphicx}
\usepackage{multirow}
\usepackage{threeparttable}
\usepackage{booktabs}
\usepackage{url}
\usepackage{geometry}
\usepackage{fancyhdr}

\fancypagestyle{firststyle}{
  \fancyhf{}
  \fancyhead[C]{2025 Asia Pacific Signal and Information Processing Association Annual Summit and Conference (APSIPA ASC)}
}

\usepackage{soul, xcolor}
\usepackage{bm}
\usepackage{amsfonts}

\begin{document}

\title{Characterization of Speech Similarity Between Australian Aboriginal and High-Resource Languages: A Case Study on Dharawal}

\author{
\authorblockN{
Ting Dang\authorrefmark{1},
Trini Manoj Jeyaseelan\authorrefmark{2}, 
Eliathamby Ambikairajah\authorrefmark{2},
Vidhyasaharan Sethu\authorrefmark{2}
}

\authorblockA{
\authorrefmark{1}
The University of Melbourne, Australia \\
E-mail: ting.dang@unimelb.edu.au}

\authorblockA{
\authorrefmark{2}
The University of New South Wales, Australia \\
E-mail: \{trini\_manoj.jeyaseelan, e.ambikairajah, v.sethu\}@unsw.edu.au}
}

\maketitle
\thispagestyle{firststyle}
\pagestyle{fancy}

\begin{abstract}
Australian Aboriginal languages are of significant cultural and linguistic value but remain severely underrepresented in modern speech AI systems. While state-of-the-art speech foundation models and automatic speech recognition excel in high-resource settings, they often struggle to generalize to low-resource languages, especially those lacking clean, annotated speech data. In this work, we collect and clean a speech dataset for Dharawal, a low-resource Australian Aboriginal language, by carefully sourcing and processing publicly available recordings. Using this dataset, we analyze the speech similarity between Dharawal and 107 high-resource languages using a pre-trained multilingual speech encoder. Our approach combines (1) misclassification rate analysis to assess language confusability, and (2) fine-grained similarity measurements using cosine similarity and Fréchet Inception Distance (FID) in the embedding space. Experimental results reveal that Dharawal shares strong speech similarity with languages such as Latin, Māori, Korean, Thai, and Welsh. These findings offer practical guidance for future transfer learning and model adaptation efforts, and underscore the importance of data collection and embedding-based analysis in supporting speech technologies for endangered language communities.
\end{abstract}

\section{Introduction}
Australia is home to a rich diversity of Aboriginal languages, many of which are critically endangered~\cite{simpson2019language}. Among these, Dharawal is one of the Indigenous languages with deep cultural significance but limited linguistic resources~\cite{capell1970aboriginal}. Despite its historical and cultural importance, Dharawal, like many other Aboriginal languages, faces challenges in preservation and technological support due to its low-resource status.

The rapid advancement of AI technologies, particularly in speech processing, has led to significant breakthroughs in automatic speech recognition (ASR)~\cite{kheddar2024automatic, chu2024qwen2}, language modeling~\cite{fathullah2024prompting}, and multilingual understanding~\cite{pratap2024scaling}. State-of-the-art models such as Whisper~\cite{radford2023robust}, Wav2Vec 2.0~\cite{baevski2020wav2vec}, and VoxLingua107 ECAPA-TDNN~\cite{chaitra2023study} have demonstrated impressive capabilities in transcribing and understanding spoken language across diverse global languages. However, these advancements predominantly focus on high-resource languages, those with abundant annotated speech data, leaving low-resource languages, including Dharawal, largely underrepresented in modern AI-driven solutions. 
Moreover, the limited availability of Dharawal speech recordings, especially the lack of transcriptions, further restricts the development of effective AI models for this language.

It is therefore essential to first collect Dharawal speech recordings for detailed analysis. While these recordings are largely unlabelled, a key initial step is to understand how Dharawal relates to high-resource languages within the multilingual speech representation space. By examining these relationships, we can more effectively leverage knowledge transfer and adaptation techniques. For instance, identifying languages with similar acoustic or phonetic characteristics enables the use of pretrained models from high-resource languages as a foundation for improving recognition or synthesis in low-resource languages. This strategy can substantially reduce dependency on large annotated low-resource languages datasets, which are often unavailable.

Existing studies have explored techniques such as transfer learning and adapters for other relatively low-resource languages~\cite{tu2019end, romanenko2022robust, hou2021exploiting, kim2025improving} such as Georgian and Kazakh, but they rely on the availability of the transcribed speech recordings from publicly accessible datasets like Common Voice~\cite{ardila2019common}.
While there are increasing studies on Australian Aboriginal languages~\cite{ambikairajah2025study}, there is currently no study on Dharawal, nor is there any publicly available labeled Dharawal speech dataset. Moreover, techniques commonly used for low-resource languages often rely on transcriptions~\cite{ahmed2021discovering} and therefore cannot be directly applied to untranscribed Dharawal speech recordings.

To bridge this gap, we first collected a Dharawal speech dataset. Further, we analyzed the speech similarity between Dharawal and other high resource languages, aiming to provide the first systematic quantification of speech similarity in the absence of transcriptions, to inform future transfer learning efforts. 
Specifically, we process the Dharawal speech recordings using pretrained multilingual encoders and quantify language similarities using two metrics: (1) the misclassification rate, which measures how frequently Dharawal is classified as other languages, and (2) cosine similarity and Fréchet Inception Distance (FID) scores~\cite{heusel2017gans}, which quantify the similarity of the speech embeddings in the shared latent speech space. While the former provides a coarse-level estimate of confusion, the latter offers a fine-grained, data-driven approach to measure affinities within the shared representational space. By quantifying these similarities, we can provide a foundation to better understand language relationships and guide the development of effective adaptation strategies for low-resource languages like Dharawal.
%Pre-trained encoders offer a powerful tool for this analysis because they abstract speech signals into fixed-dimensional embeddings capturing rich phonetic and linguistic features. Since these models are trained on large and diverse datasets, they provide a meaningful framework to compare languages even with minimal or no additional training on low-resource languages. This capability is essential for languages like Dharawal, where conventional data-driven methods struggle.
%By leveraging these techniques, our study aims to explore the application of multilingual pre-trained encoders to analyze and understand Dharawal speech within a broader context. By leveraging a large-scale language identification (LID) model, we seek to quantify Dharawal's similarities to high-resource languages and provide insights into its characteristics. Our approach not only helps in identifying cross-linguistic relationships but also contributes to the broader goal of digital preservation and technological inclusion of Australian Aboriginal languages.
The contributions of this paper are summarized as follows:
\begin{itemize}
    \item We collected and cleaned the first speech dataset for the Australian Aboriginal language, Dharawal.
    \item We proposed the first speech embedding analysis for Dharawal by examining its misclassification rate and embedding similarity using a pretrained multilingual encoder, comparing it with other high-resource languages.
    \item Experimental results show that Dharawal is most similar to Latin, Māori, Korean, Thai, and Welsh. Although the ranking order differs slightly between misclassification rate and embedding similarity, the top 10 most similar languages largely overlap.
\end{itemize}
\section{Related Work}
\subsection{Dharawal}
Dharawal is an Australian Aboriginal language spoken by people in the region surrounding modern day Sydney, New South Wales~\cite{timberydharawal}. Like most Indigenous Australian languages, Dharawal faces challenges due to language shift and lack of intergenerational language transmission. Although efforts to revitalize and preserve the language are ongoing, available linguistic resources remain sparse. %Most existing documentation and resources has been collected through community effort and archival recordings~\td{ref}. 
To the best of our knowledge, no prior work has collected the Dharawal speech recordings or analyzed speech embedding similarities using pre-trained speech encoders for Dharawal.

\subsection{Language similarity identification}
%Language similarity detection in speech is a growing field, particularly within multilingual and low-resource language research. 
Language similarity detection plays a critical role in multilingual speech and language processing, particularly for low-resource languages. By identifying high-resource languages that are acoustically or structurally similar, researchers can leverage transfer learning~\cite{yi2018language, byambadorj2021text}, cross-lingual adaptation~\cite{tu2019end}, and multilingual modeling~\cite{toshniwal2018multilingual, pratap2024scaling} more effectively.
Prior work has utilized phoneme-based comparisons \cite{ingram2024cross}, levering the lingusitic similarities. % acoustic feature analysis \cite{article}, and more recently, deep learning–based embeddings \cite{8461375, zhang15c_interspeech}.
% While most deep-learning based methods rely on supervised training or fine-tuning using labeled datasets, our approach focuses more on unsupervised analysis of language similarity using pre-trained embeddings without retraining \cite{baevski2020wav2vec}. 
However, the exploration of speech similarity beyond purely linguistic content remains limited. With the advancement of pretrained speech encoders, such as Wav2Vec 2.0~\cite{baevski2020wav2vec}, which are trained on diverse languages and speech recordings, and the growing availability of large curated multilingual speech datasets like VoxLingua107~\cite{valk2021slt}, it is now increasingly feasible to perform speech similarity identification without the need for extensive annotated data.
% Although the existing deep learning approaches share the commonality with our proposed approach, it differs in two different ways. One method is to use the embeddings directly from a pre-trained ECAPA-TDNN model trained on the VoxLingua107 dataset without any task specific fine-tuning. The other method is the application of unsupervised similarity analysis using cosine similarity and Frechet Inception Distance between languages to identify similar languages. 

% and the more and more curated large multilingua lspeech dataset such as VoxLingua107 dataset~\cite{valk2021slt} have enabled cross-lingual comparisons without requiring large annotated datasets. 

%Studies have used cosine similarity between speech embeddings, and Frechet Inception Distance (FID) to measure similarity across languages. These methods help identify which high-resource languages can act as a foundation, which makes it easier to adapt speech models for less documented languages. 
\section{Dataset}
 The Dharawal speech dataset used in this study was derived from publicly available materials on the Dharawal Words website\footnote{\url{https://www.dharawalwords.com.au}}, which shares language resources from the native Dharawal speaker for educational and cultural preservation purposes. We collected the recordings following strict ethical guidelines, and the resulting dataset remains private and is used solely for research. The dataset included 475 utterances.

All recordings were manually reviewed to ensure clarity and audio quality. Specifically, each file was inspected to identify and remove segments containing excessive noise, corruption, or irrelevant content. Recordings with unclear speech or extended silence were excluded to maintain data integrity. This cleaning step helps minimize unnecessary padding and ensures that the resulting embeddings capture meaningful linguistic features rather than background noise.

The Dharawal speech segments were generally short and primarily consisted of individual words, with only a small number of utterances containing sentence-level content. To generate more meaningful samples for analysis, shorter recordings were concatenated in a way that preserved contextual coherence. This process was conducted using Audacity~\cite{audacity2017audacity}, an open-source audio editing tool, and included trimming silences at the beginning and end of each clip before assembling them into longer sequences. The resulting dataset comprises utterances approximately 10 to 15 seconds in length.

All audio files were saved in WAV format at a 16 kHz sampling rate. No data augmentation was applied in order to preserve the natural acoustic characteristics of the original recordings.

\begin{figure*}[t]
    \centering
    \includegraphics[width=0.95\linewidth]{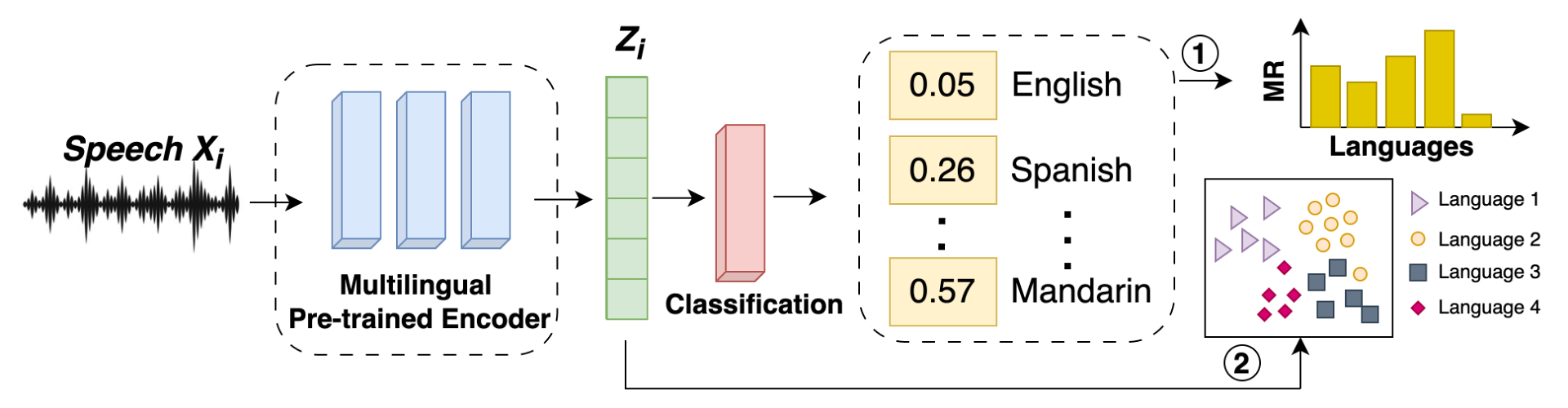}
    \caption{System Overview. The input speech utterance in Dharawal, denoted as $\bm{x}_i$, is processed by a multilingual pre-trained encoder to obtain embeddings $\bm{z}_i$. These embeddings are then passed through a classification head to produce the final predicted language probabilities. We propose two approaches: (i) estimating the misclassification rate (MR) based on the probability outputs, and (ii) quantifying embedding similarity and FID scores in the latent space.}
    \label{fig:overview}
\end{figure*}

\section{Methods}
\subsection{Overview}
As shown in Figure~\ref{fig:overview}, given a speech utterance $\bm x_i$ from our Dharawal dataset $\mathcal{D}$, we process the input through a pretrained multilingual encoder $f(\cdot)$ to obtain speech embeddings $\bm z_i$ . This enables us to perform a two-fold analysis: (i) estimating the misclassification rate by evaluating how often the embeddings of $\bm z_i$ are incorrectly classified as belonging to other high-resource languages, providing insight into the encoder’s discrimination capability for low-resource Aboriginal languages; and (ii) quantifying the embedding similarity between Dharawal and other languages using cosine similarity and FID scores, where cosine similarity captures the directional alignment between embedding vectors, and FID quantifies the overall statistical distance between their distributional properties (mean and covariance). This framework allows us to assess both the coarse-level separability and the fine-grained relative closeness of Dharawal embeddings within the multilingual embedding space.

\subsection{Misclassification Rate Analysis for Language Identification}
To identify high-resource languages similar to Dharawal, we employed a multilingual pretrained encoder that includes a wide variety of languages for analysis. Specifically, we used the state-of-the-art VoxLingua107-ECAPA-TDNN model~\cite{valk2021slt}, as they have been pretrained on a diverse corpus encompassing 107 distinct languages, ensuring broad linguistic representation across multiple language families. When presented with a Dharawal speech utterance $\bm x_i$, the model generates a probability distribution over the 107 language classes $\hat{y_i}$. The predicted label corresponds to the language with the highest assigned probability. The misclassification rate is determined by measuring the frequency with which Dharawal are incorrectly assigned to any of the known language categories as:
\begin{equation}
  \text{MR} = \frac{1}{N} \sum_{i=1}^N \mathbb{I} \big( \hat{y}_i \neq \text{Dharawal} \big)  
\end{equation}
where $N$ is the total number of Dharawal utterances, $\hat{y}_i$ is the predicted label for the $i$-th utterance, and $\mathbb{I}(\cdot)$ is the indicator function that equals 1 if the condition is true, and 0 otherwise. A higher misclassification rate indicates greater similarity between Dharawal and the corresponding language, and vice versa.

\subsection{Quantification of Embedding Similarity}
To evaluate the embedding-level similarity between Dharawal and other high-resource languages, we employ cosine similarity and FID scores. These metrics quantify the difference between the speech embeddings of Dharawal and a target language in a shared embedding space.

\paragraph{Embedding Extraction}
Speech embeddings are obtained using the encoder part of the pre-trained multilingual VoxLingua107-ECAPA-TDNN model~\cite{valk2021slt}. Given a speech utterance $\bm x_i$, the encoder maps it into a fixed-dimensional embedding vector:
\begin{equation}
 \bm{z}_i = f(\bm x_i)  
\end{equation}
where $\bm{z}_i$ is the resulting speech embedding in a high-dimensional latent space as shown in Figure~\ref{fig:overview}. 

\paragraph{Cosine Similarity.}
For cosine similarity, the model pre-stores class centroids for the pretrained 107 high-resource languages, where $\bm \mu_L$ denotes the centroid, i.e., the mean embedding vector of the $L^{th}$ target language. We extract the centroid embedding to represent each high-resource language in the shared latent space. For the Dharawal input utterances ${\bm {x}_i}$, we compute their corresponding embeddings and calculate the mean to obtain the centroid representation $\bm \mu_{Dharawal}$ as:
\begin{equation}
    \bm \mu_{Dharawal} = \frac{1}{N}\sum_{i=1}^{N} \bm z_i
\end{equation}

The cosine similarity between Dharawal and a target language $L$ is then computed as:
\begin{equation}
 % cos(L) = \frac{1}{N} \sum_{i=1}^{N} \frac{\boldsymbol{x}_i \cdot \boldsymbol{\mu}_L}{\|\boldsymbol{x}_i\| \, \|\boldsymbol{\mu}_L\|} 
 \text{CosSim}(\boldsymbol{\mu}_\text{Dharawal}, \boldsymbol{\mu}_L) = \frac{\boldsymbol{\mu}_\text{Dharawal} \cdot \boldsymbol{\mu}_L}{\|\boldsymbol{\mu}_\text{Dharawal}\| \|\boldsymbol{\mu}_L\|}
\end{equation}
A higher cosine similarity score indicates that the embeddings of Dharawal are statistically closer to those of the target language, suggesting stronger linguistic or acoustic similarities. Conversely, a lower cosine similarity implies greater divergence in the representation space.
% where $N$ is the total number of Dharawal speech embeddings and $\boldsymbol{\mu}_L$ is the centroid of the $L^{\text{th}}$ high-resource language.

% \paragraph{Quantifying Embedding Similarity.}
% To measure the similarity between the two languages, we compute the cosine similairty and FID based on the statistical properties of their embedding distributions. 
\paragraph{FID scores.}
% For each language, we collect a set of embeddings: $\mathcal{Z}_D = \{\bm{z}_i^D\}_{i=1}^{N_D}$ for Dharawal and $\mathcal{Z}_L = \{\bm{z}_i^L\}_{i=1}^{N_L}$ for a target language, where $N_D$ and $N_L$ represent the number of samples for Dharawal and the target language, respectively.

Regarding FID scores, the computation requires estimating the distributional properties of the embedding space for each language, specifically the mean and covariance of the embeddings. Therefore, we need to collect speech recordings for each high-resource languages to obtain representative embedding clusters. For comparative analysis, we select the top 10 most similar high-resource languages%and the top 10 least similar languages
, based on prior misclassifications rates, as the target languages for FID evaluation. The final dataset for FID score analysis includes the 10 selected high-resource languages, alongside the Dharawal speech recordings.

%we specifically collect a subset of speech recordings for each language and extract their corresponding speech embeddings. 
Specifically, let $(\mu_{Dharawal}, \Sigma_{Dharawal})$ and $(\mu_L, \Sigma_L)$ be the mean vectors and covariance matrices of the embeddings for Dharawal and the target language, respectively. The FID score are then calculated as follows:
\begin{equation}
\small
    \operatorname{FID}(\mathcal{Z}_D, \mathcal{Z}_L) = || \mu_D - \mu_L ||^2 + \operatorname{Tr}\left(\Sigma_D + \Sigma_L - 2(\Sigma_D \Sigma_L)^{\frac{1}{2}}\right)
\end{equation}
where $|| \cdot ||^2$ denotes the squared Euclidean distance, and $\operatorname{Tr}(\cdot)$ represents the matrix trace operation.
A lower FID score indicates that the embeddings of Dharawal are statistically closer to those of the target language, and vice versa.

% \paragraph{Target Language Selection}
% To perform comparative analysis, we selected two groups of high-resource languages from the VoxLingua107 dataset.

% The top 10 most similar languages were identified based on preliminary language identification results using the VoxLingua107-ECAPA-TDNN model. Languages with the highest classification probabilities for Dharawal utterances were considered to be the most similar.

% Conversely, the top 10 least similar languages with the lowest classification probabilities and minimal acoustic resemblance were selected for contrastive analysis.

\section{Experimental Setup}
% \subsection{Dataset}
% \td{Trini, write the dataset you have for now, e.g, how long (duration), how many speakers, how many utternces, etc.}

\begin{figure*}[t]
    \centering
    \includegraphics[width=1\linewidth]{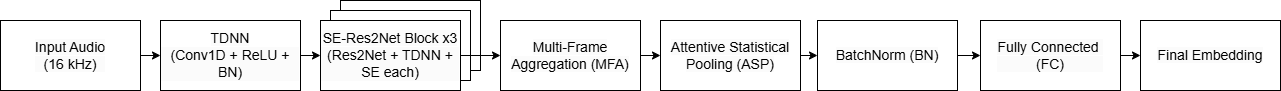}
    \caption{Multilingual pre-trained encoder architecture for similarity characterization.}
    \label{fig:encoder}
\end{figure*}

\subsection{Multilinguial pre-trained encoder}
The multilingual pre-trained encoder VoxLingua107-EPACA-TDNN is built on the ECAPA-TDNN (Emphasized Channel Attention, Propagation, and Aggregation Time Delay Neural Network) architecture~\cite{he2024hierarchical}, as shown in Figure~\ref{fig:encoder}. It was trained on the VoxLingua107 dataset~\cite{valk2021slt}, which consists of over 6,000 hours of speech from 107 different languages. Specifically, it first extracts frame-level features from raw audio inputs using a series of 1D convolutional layers, which capture local acoustic patterns and temporal dependencies directly from the waveform. To enhance feature representation, Squeeze-Excitation (SE) blocks are employed, dynamically recalibrating channel-wise responses to emphasize informative channels. Temporal processing is managed by TDNN layers, which apply 1D convolutions over varying contexts to incorporate long-range dependencies. An attention mechanism then aggregates frame-level features, weighting each frame's contribution based on its importance, and outputs a 256-dimensional embedding vector that encapsulates the global linguistic and acoustic characteristics of the speech segment, optimized for language classification and similarity analysis.%\td{Trini, could you please draw a brief figure for the model structure? }

% TalTechNLP’s TDNN model\td{ref} is a less intensive and faster alternative to the VoxLingua107-EPACA-TDNN model. It is more suited for real-time language identification. This model uses
% frame-wise feature extraction instead of channel attention mechanisms.

\subsection{Implementation details}
%Two pre-trained models were used for language identification - (1) TalTechNLP/voxlingua107-EPACA-TDNN which outputs the cosine similarity scores for the 107 languages present in the VoxLingua107 dataset and (2) speechbrain/lang-id-voxlingua107-ecapa which outputs the log-softmax probabilities for the same 107 languages. 62 cleaned up Dharawal utterances were used as input for both models.

During the inference phase, the multilingual encoder processes input Dharawal utterances to evaluate (i) the misclassification rate and (ii) embedding similarity. For the misclassification rate, we computed the error using the softmax output probabilities. For cosine similarity and FID scores, we extracted the 256-dimensional embeddings for each of the 107 pretrained languages and compared it to the embeddings of Dharawal to quantify similarity. To ensure a fair comparison, the number of utterances used for FID computation was kept consistent across all languages.

% To evaluate language similarity, the encode\_batch method of the ECAPA-TDNN model was used to extract the embeddings of the audio files. The mean embedding of each language is extracted and then compared with the mean embedding of Dharawal for the cosine similarity analysis. Similarly, the Frechet Inception Distance (FID) scores were also calculated using the means and covariances of embeddings. 

% To compute misclassification rates, the cosine similarity output from the TalTechNLP model and the softmax probability from the SpeechBrain model output were used to identify the top predicted language for each of the Dharawal utterances. These were then tallied and normalized over the total number of Dharawal utterances (62). All preprocessing was done using standard packages and libraries, including NumPy, SciPy, and Matplotlib.

All experiments were conducted using Python 3.11.12 with PyTorch and the SpeechBrain libraries. All computations were performed on Google Colab using CPU resources.

\begin{figure}[t]
    \centering
    \includegraphics[width=1\linewidth]{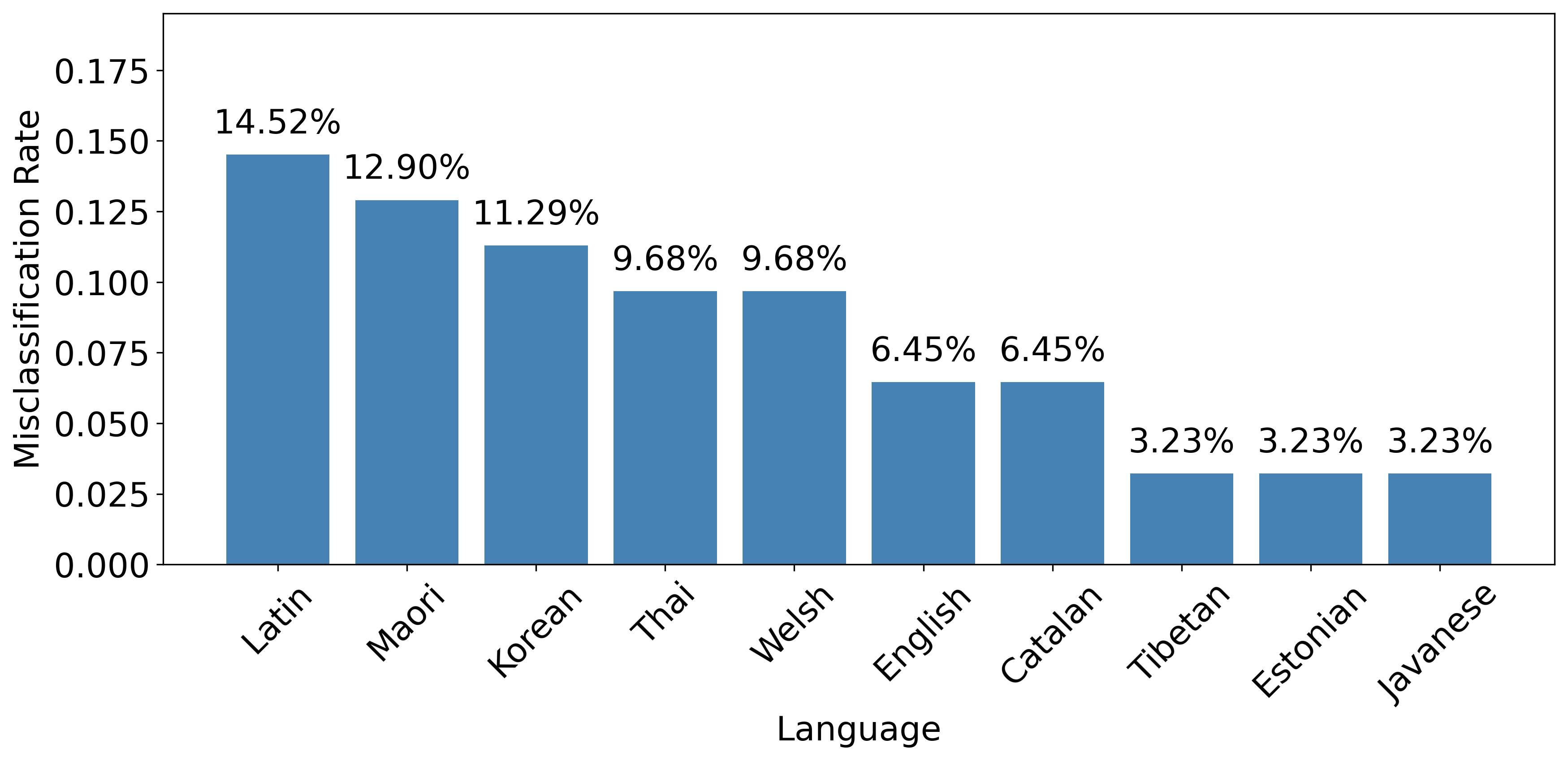}
    \caption{Misclassification rates for the top 10 languages identified as similar to Dharawal using VoxLingua107-EPACA-TDNN, respectively.}
    \label{fig:1}
\end{figure}

\section{Results}
\subsection{Misclassficaition rate}
The misclassifications rate for the pretrained encoder are shown in Figure~\ref{fig:1}. It displays the top 10 languages with the highest misclassification rates for Dharawal.
It is evident that Dharahal is most frequently misclassified as Latin, with a misclassification rate of 14.52\%, indicating a notable degree of similarity between the two. Classical Latin has a relatively simple and consistent vowel system and syllable structure (often Consonant-Vowel or Consonant-Vowel-Consonant)~\cite{cser2016aspects}, characteristics that are also common in many Australian Aboriginal languages. Furthermore, both may lack certain consonant clusters or tonal variation, making them acoustically similar at a superficial level. This is followed by Māori, Korean, Thai, and Welsh, which also exhibit relatively high confusion rates. The confusion with Māori is more expected. Both languages are indigenous to the southern hemisphere and are known for having small phoneme inventories, frequent use of open syllables (Consonant-Vowel patterns), and clear, rhythmic prosody~\cite{harlow2007maori}. These shared features can easily cause acoustic overlap in models that rely on phonetic features alone. These similarities suggest opportunities to leverage the acoustic patterns of similar languages for data augmentation or transfer learning, to further improve AI models for under-resourced Dharawal languages.

\begin{table*}[ht]
    \centering
    \caption{Cosine similarity between Dharabal speech embeddings and speech embeddings from other 10 most similar high-resource languages.}
    \begin{tabular}{c|c|c|c|c|c|c|c|c|c|c|c|c}
    \toprule
         Languages & Thai&Welsh&Korean&Tibetan&Estonian&English&Javanese&Catalan&Māori&Latin   \\
         \midrule
         Cos similarity& 0.9078&0.8969&0.8937&0.8845&0.8785&0.8749&0.8743&0.8562&0.8533&0.8238  \\
    \bottomrule
    \end{tabular}
    \label{tab:1}
\end{table*}

\subsection{Embedding similarity}
To provide a more fine-grained analysis of embedding similarities, we report cosine similarity and FID scores between the encoders and the top 10 most similar high-resource languages. %as well as the 10 least similar languages.  

\begin{figure}[t]
    \centering
    \includegraphics[width=1\linewidth]{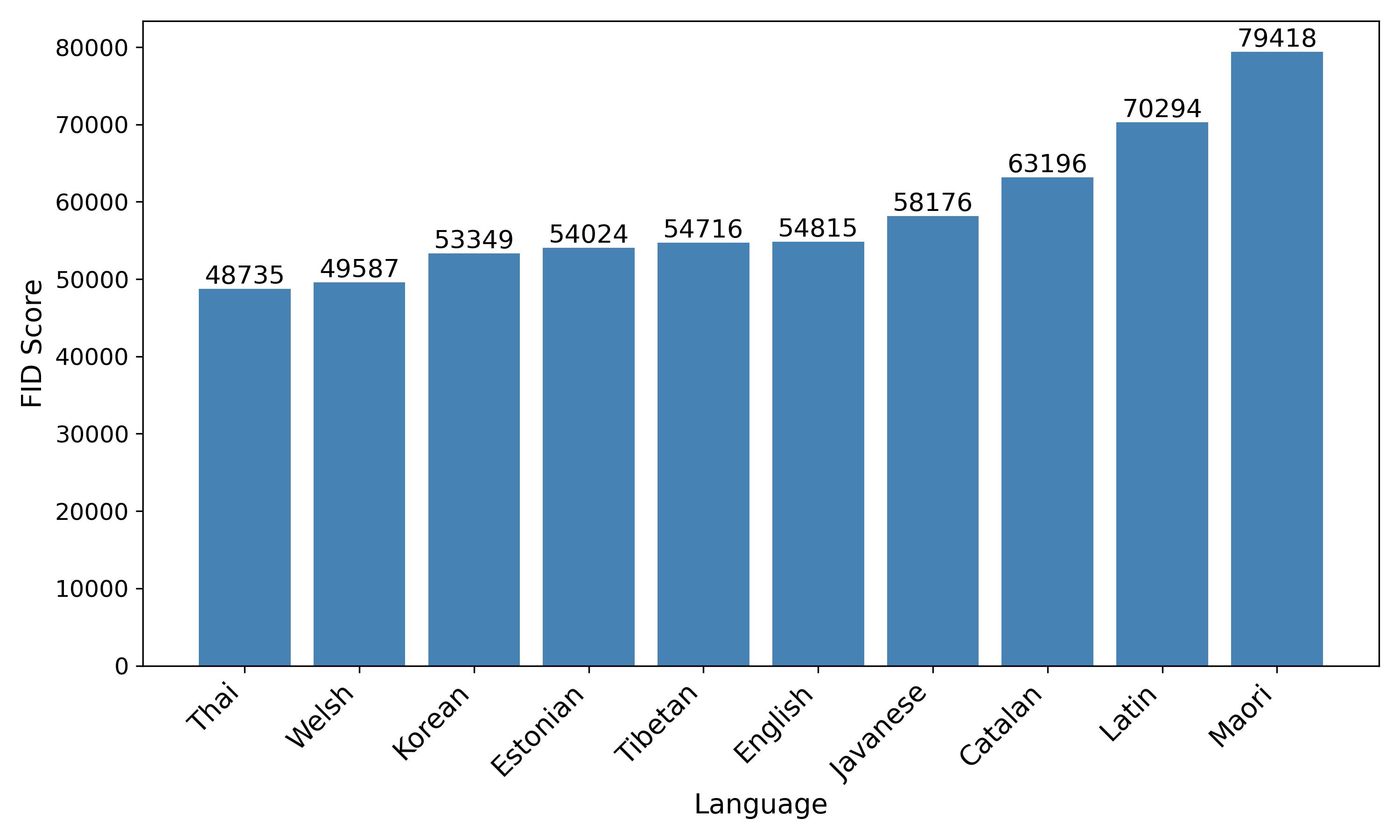}
    \caption{FID scores between Dharawal speech embeddings and the speech embeddings of the 10 most similar languages.}
    \label{fig:fid}
\end{figure}

For cosine similarity and FID scores of the top 10 most similar languages, as shown in Table~\ref{tab:1} and Figure~\ref{fig:fid}, the same set of languages appears with nearly identical ranking orders. However, these fine-grained similarity measures show slight differences in ranking when compared to the misclassification rates. Notably, the embeddings show the highest cosine similarity and FID scores for Thai, Welsh, and Korean, while Latin and Māori which are ranked highest in the misclassification rate are ranked least using the cosine similarity and FID scores. This discrepancy arises because cosine similarity, FID score and misclassification rate capture fundamentally different aspects of language similarity and model behavior. Cosine similarity and FID score measure the geometric closeness of language embeddings in the learned feature space, reflecting intrinsic representational similarity based on shared phonetic, phonological, or acoustic patterns. Thus, languages like Thai, Welsh, and Korean may cluster closely due to underlying phonetic traits that the model encodes similarly. In contrast, the misclassification rate reflects practical classification challenges and model errors when distinguishing between languages. The high misclassification rates for Latin and Māori suggest that the model struggles to distinguish these languages from others. This may be due to partially overlapping acoustic features, insufficient training data, or noise and variability within the Māori recordings. Despite their relatively low overall cosine similarity, the distribution or clustering of these languages may contain subsets that overlap significantly with others, leading to confusion during classification. % the classifier might confuse these languages with others because the model’s decision boundaries are influenced by factors beyond pure embedding similarity. 
These metrics complement each other, offering a comprehensive picture of language similarity and model performance.

% For the 10 least similar languages, as shown in Table~\ref{tab:2}, we observe \td{to continue.}

% \begin{table*}[ht]
%     \centering
%     \caption{Cosine similarity between Dharabal speech embeddings and speech embeddings from other 10 least similar high-resource languages.}
%     \begin{tabular}{c|c|c|c|c|c|c|c|c|c|c|c|c}
%     \toprule
%          Languages & Esperanto&Welsh&Thai&Māori&&&&&   \\
%          \midrule
%          Cos similarity& 0.7054&0.6996&0.6828&0.6763&&&&&  \\
%     \bottomrule
%     \end{tabular}
%     \label{tab:2}
% \end{table*}

% \begin{table*}[ht]
%     \centering
%     \caption{FID score between Dharawal speech embeddings and the speech embeddings of the 10 most similar languages.\td{convert it to a bar plot?}}
%     \begin{tabular}{c|c|c|c|c|c|c|c|c|c|c|c|c}
%     \toprule
%          Languages &Thai&Welsh&Korean&Estonian&Tibetan&English&Javanese&Catalan&Latin&Māori   \\
%          \midrule
%          FID&48,735&49,587&53,349&54,024&54,716&54,815&58,176&63,196&70,294&79,418  \\
%     \bottomrule
%     \end{tabular}
%     \label{tab:3}
% \end{table*}

\subsection{Visualization}
\begin{figure}[t!]
    \centering
    \includegraphics[width=1\linewidth]{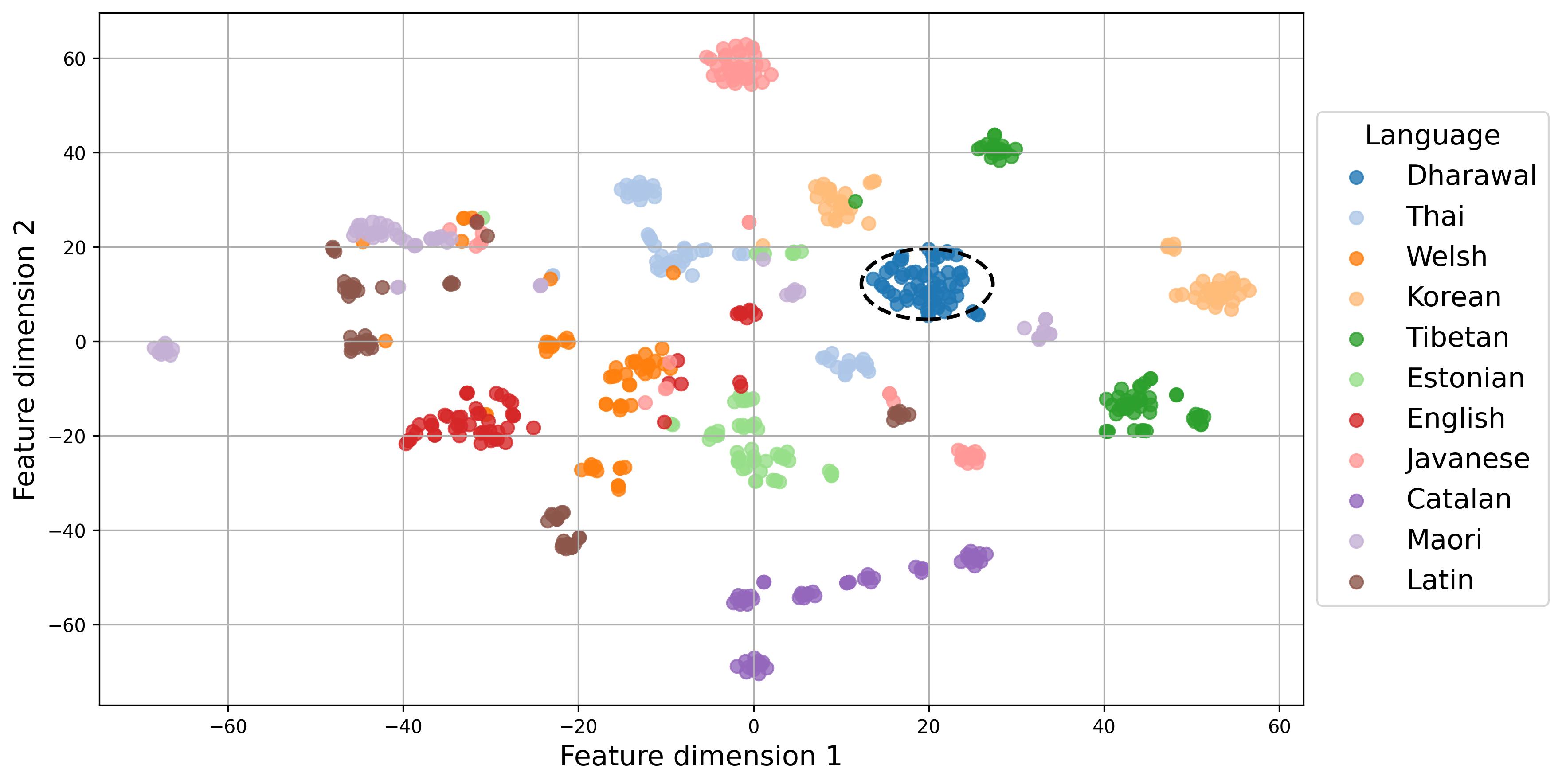}
    \caption{t-SNE plot of the embeddings for Dharawal and other high-resource languages.}%\td{@Trini, could you update the plot by renaming the x- and y-axes to "Feature Dimension 1" and "Feature Dimension 2"? Please circle Dharawal to make it clearly identifiable. Also make Dharawal the first row in the legend, and the order same as cosine similairty top 10 lanagues. Make sure the legend, axis labels, and font sizes are clearly visible and easy to read. Additionally, try adding a grid to see if it improves the overall clarity of the visualization. Don't use the outer rectangular box. }}
    \label{fig:2}
\end{figure}

To further explore the embedding space, we visualize the latent representations of the top 10 most similar languages with respect to Dharawal. The t-SNE (t-distributed Stochastic Neighbor Embedding) plot in Figure~\ref{fig:2} demonstrates patterns that largely align with the computed similarity metrics. Dharawal exhibits the closest proximity to Thai (cyan) and and Korean (light orange) 
in the latent space, consistent with their high computed similarity scores. %Although Welsh (green) with sparse points spreading across the latent space does not show low distance from the visualization perspective, the main cluster near the Dharawal cluster still leads to the centroid of Welsh close to Dharawal, therefore ranks second in computed similarity.

For Javanese (pink) and Estonian (light green), the main cluster position closely to Dharawal, but scattered outlier points distributed across the latent space, likely reflecting significant internal language variability, cause the centroid to shift away from Dharawal, resulting in moderate similarity ranking among the ten languages. Catalan (purple) is positioned far from Dharawal, therefore ranked the 8th in the top 10 languages. Notably, small clusters of Māori (light purple) appear on the right side of the Dharawal clusters, in close proximity to them, which may explain the observed high misclassification rates between these languages. However, sparse Māori data points and large clusters extending toward the left side of the embedding space pull the Māori centroid away from Dharawal, consequently reducing their computed similarity score. These observations suggest that Māori and Dharawal share substantial underlying similarity in the latent representation in general.

% Interestingly, while Latin (brown) demonstrates the highest confusion rates in classification tasks, its cluster appears distant from Dharawal in the latent space. This discrepancy between classification confusion and latent space distance may be attributed to overlapping decision boundaries in specific feature dimensions that are not captured in the two-dimensional t-SNE projection.

%languages with high cosine similarity, such as Thai, Welsh, and Korean, tend to form tighter and more overlapping clusters, indicating shared structure in their representations. In contrast, languages like Latin and Māori are more distinctly separated in the latent space, aligning with their lower cosine similarity scores. The updated figure includes clearer axis labels ("Feature Dimension 1" and "Feature Dimension 2") and highlights Dharawal with a visual marker for easier identification. This visualization reinforces the patterns observed in the similarity metrics and provides an intuitive view of how language representations are organized within the model.

\section{Conclusion}
In this work, we collect and clean a dataset for the Dharawal language, contributing a valuable resource to the growing body of speech data for Australian Aboriginal languages. We investigated two approaches for analyzing speech similarity using pre-trained multilingual speech encoders. Specifically, we evaluated classification separability via misclassification rates and quantified embedding similarity through cosine similarity and Fréchet Inception Distance scores.

Our results consistently indicate that Dharawal exhibits strong acoustic similarity to high-resource languages such as Latin, Māori, Korean, Thai, and Welsh. This finding suggests that speech models trained on these languages may serve as effective bases for transfer learning or adaptation to support Dharawal and potentially other low-resource Aboriginal languages—without requiring large-scale transcribed data.

This study underscores the importance of systematic similarity analysis for informing downstream development of speech technologies for endangered languages. Future work could expand on these insights by exploring multiple model architectures and larger datasets as they become available, developing lightweight adaptation techniques based on the most acoustically similar high-resource languages, and applying this framework to additional Indigenous Australian languages to further support linguistic preservation and accessibility.

% Our findings revealed the high similarity between Dharawal and Latin, Māori, Korea, Thai and Māori , in both classification and embedding space structure. This suggests that Dharawal shares acoustic characteristics with certain high-resourced languages, therefore a potential pathways for transfer learning or adaptation of existing speech models for these lanagues to under-resourced Aboriginal languages.

% Overall, our study highlights the importance of both data curation and embedding analysis in understanding and supporting speech technologies for endangered languages. Future work could explore fine-tuning strategies based on the identified similarities, as well as extend this framework to additional Indigenous languages in Australia and beyond.

\section*{Acknowledgment}
The authors would like to thank the School of Electrical Engineering and Telecommunications at UNSW Sydney, Australia, for providing funding for this research initiative.

% \begin{thebibliography}{1}

% \bibitem{1}
% G.~Eason, B.~Noble, and I.~N.~Sneddon, ``On certain integrals of
% Lipschitz-Hankel type involving products of Bessel functions,''
% \emph{Phil. Trans. Roy. Soc. London,} vol. A247, pp. 529-551, April
% 1955.

% \bibitem{2}
% J.~Clerk~Maxwell, \emph{A Treatise on Electricity and Magnetism,}
% 3$^{\rm rd}$ ed., vol. 2. Oxford: Clarendon, 1892, pp.68-73.

% \bibitem{3}
% I.~S.~Jacobs and C.~P.~Bean, ``Fine particles, thin films and exchange
% anisotropy,'' in \emph{Magnetism,} vol. III, G.T. Rado and H. Suhl,
% Eds. New York: Academic, 1963, pp. 271-350.

% \bibitem{4}
% K.~Elissa, ``Title of paper if known,'' unpublished.

% \bibitem{5}
% R.~Nicole, ``Title of paper with only first word capitalized,''
% \emph{J. Name Stand. Abbrev.,} in press.

% \bibitem{6}
% Y.~Yorozu, M.~Hirano, K.~Oka, and Y.~Tagawa, ``Electron spectroscopy
% studies on magneto-optical media and plastic substrate interface,''
% \emph{APSIPA Transl. J. Magn. Japan,} vol. 2, pp. 740-741, August 1987
% [\emph{Digests 9$^{\rm th}$ Annual Conf. Magnetics Japan,} p. 301,
% 1982].

% \bibitem{7}
% M.~Young, \emph{The Technical Writer's Handbook.} Mill Valley, CA:
% University Science, 1989.

% \end{thebibliography}

% \printbibliography
\bibliographystyle{IEEEtran}
\bibliography{Latex-2025/mybib}

\end{document}